\documentclass[twocolumn,superscriptaddress,aps,pra]{revtex4-2}

\usepackage{amsmath,amssymb,amsthm}
\usepackage{physics}
\usepackage{amsfonts}
\usepackage{mathrsfs}
\usepackage{graphicx}
\usepackage{tabularx}
\usepackage{enumerate}
\usepackage{dcolumn}
\usepackage{bm}
\usepackage{xcolor}
\usepackage[normalem]{ulem}
\usepackage[colorlinks,linkcolor=blue,citecolor=blue,urlcolor=blue]{hyperref}

\usepackage{braket}
\usepackage{braket}

\begin{document}
	\preprint{APS/123-QED}
	\global\long\def\id{\mathbbm{1}}
	\global\long\def\ui{\mathbbm{i}}
	\global\long\def\ud{\mathrm{d}}
	
	\title{Dissipative charging of tight-binding quantum batteries}
	
	\author{Mingdi Xu}
	\affiliation{School of Physics, Nankai University, Tianjin 300071, China}
	
	\author{Yiming Liu}
	\affiliation{School of Physics, Nankai University, Tianjin 300071, China}
	
	\author{Yefeng Song}
	\affiliation{School of Physics, Nankai University, Tianjin 300071, China}
	
	\author{Xiang-Ping Jiang}
	\email{2015iopjxp@gmail.com}
	\affiliation{School of Physics, Hangzhou Normal University, Hangzhou, Zhejiang 311121, China}
	
	\author{Lei Pan}%
	\email{panlei@nankai.edu.cn}
	\affiliation{School of Physics, Nankai University, Tianjin 300071, China}
	
	\begin{abstract}
		We investigate autonomous dissipative charging mechanisms for lattice quantum batteries within the framework of open quantum systems. Focusing on engineered Markovian dissipation, we show that appropriately designed Lindblad jump operators can drive tight-binding systems into highly excited band-edge states, resulting in steady states with large ergotropy. We illustrate this mechanism in a one-dimensional tight-binding chain and in a two-dimensional graphene lattice. We find that disorder enhances the charging power, indicating that dissipation-assisted localization effects can be beneficial for energy storage. Moreover, the dissipative charging process remains robust against additional local dephasing noise. Our results establish bond dissipation as an effective and physically transparent mechanism for charging lattice quantum batteries in realistic open-system settings.
	\end{abstract}
	
	\maketitle
	
	\section{Introduction}
	
	Quantum batteries, finite-dimensional quantum systems designed to store and deliver useful work, provide a concrete setting to explore fundamental questions at the interface of thermodynamics and quantum physics \cite{Fannes2013quantumbatteries,binder2015quantacell,campaioli2017enhancing,campaioli2024colloquium}. In this context, the performance of a quantum battery is naturally characterized by its ergotropy, defined as the maximum work extractable from a quantum state via unitary operations with respect to a fixed system Hamiltonian~\cite{Farina2018ChargerMediated,Carrega2020Dissipative,Santos2020SelfDischarging,Zakavati2020BoundsPower,Tirone2023NoisyBatteriesCoherence}. States with zero ergotropy are passive, while non-passive states encode useful nonequilibrium resources that can be converted into work. A central goal is to identify mechanisms that generate and stabilize highly non-passive states with large ergotropy.

A large body of work has shown that coherence, correlations, and collective effects can significantly enhance the charging power and extractable work in closed many-body systems~\cite{Mayo2022CollectiveDissipative,Carrasco2021CollectiveEnhancement,Cakmak2020ErgotropyCoherences,Ahmadi2025HarnessingNoise,Canzio2024SingleAtomDissipation,Quach2020OrganicQB,Pokhrel2024LargeCollectivePower,Salvia2022RepeatedInteractions,Gyhm2024Beneficial,Song2026QuantumSteering,Zhang2024QuantumBattery,Song2022EnvironmentMediated,Song2024EntropicUncertainty,Aziz2025SpectralBounds}. In realistic settings, however, quantum batteries operate as open quantum systems subject to decoherence and dissipation. Standard Markovian thermalization or amplitude-damping processes typically drive the system toward Gibbs states or the ground state, which are completely passive and thus possess vanishing ergotropy for a single reservoir~\cite{Carrega2020Dissipative,Santos2020SelfDischarging,Zakavati2020BoundsPower,Tirone2023NoisyBatteriesCoherence}. This leads to self-discharging and aging effects, observed both in theoretical models and in experimental platforms such as superconducting circuits and NV centers~\cite{Pirmoradian2019Aging,Song2025SelfDischargingMitigated,Hu2021SuperconductingQB,Malavazi2025WeakMeasurement}. A key challenge is therefore to devise charging protocols that remain effective in the presence of unavoidable environmental noise.
	
	Dissipation, however, can also be engineered as a resource. Recent studies have demonstrated that suitably engineered environments can be harnessed for quantum batteries. Non-Markovian reservoirs or structured spectral densities can slow down energy leakage or induce partial revivals~\cite{Kamin2019NonMarkovian,Tabesh2020EnvironmentMediated,Bhanja2023QBM,Shen2022NonMarkovianReservoirs,Zhao2025NonMarkovianSpinChain,Morrone2022CollisionalNonMarkovian,Song2023RemoteCharging,Tirone2024WaveguideQEDStorage,Xu2024InhibitingSelfDischarge,Hadipour2024CavityBased,Sen2023NoisyBatteries,Cavaliere2025DynamicalBlockade}. Moreover, following the seminal proposal of dissipative charging in Ref.~\cite{Barra2019DissipativeCharging}, autonomous dissipative charging schemes have been proposed in which engineered reservoirs drive the battery toward non-passive steady-states with finite ergotropy, without explicit time-dependent control on the battery degrees of freedom~\cite{Mayo2022CollectiveDissipative,Carrasco2021CollectiveEnhancement,Cakmak2020ErgotropyCoherences,Ahmadi2025HarnessingNoise,Hadipour2025NonequilibriumBathModulation,Khoudiri2025CorrelationEnhanced,Ahmadi2024Superoptimal,Zhang2025DissipativeQutrit,Ferraro2026Opportunities}. Related driven-dissipative schemes, where coherent driving and engineered dissipation act jointly, have also been shown to enhance power and ergotropy~\cite{Wang2022DrivenDissipative,Pokhrel2024LargeCollectivePower}. These works establish reservoir engineering as a powerful tool for mitigating dissipation-induced degradation.
	
	Within this broad landscape, lattice and spin-chain quantum batteries offer a particularly appealing platform. Tight-binding and spin-chain Hamiltonians arise naturally in condensed-matter systems and quantum simulators, and their band structures provide additional degrees of freedom for energy storage. Previous studies have shown that interactions, hopping, topology, or feedback can enhance stored energy or stabilize charged states under dissipation~\cite{Zhao2020InteractingSpins,Zhao2025NonMarkovianSpinChain,Sun2024CavityHeisenberg,Yao2022FeedbackSpinChain,Yadav2025CollectiveQBs,Zhao2025RapidStable,Fajar2025ZenoLike,Zhao2025CatalyticNZ,Hadipour2025NonequilibriumBathModulation,Lu2024TopologicalQB,Song2023RemoteCharging,Tirone2024WaveguideQEDStorage,Ahuja2025CorrelatedReservoirs}. In most of these proposals, however, dissipation primarily plays a stabilizing or protective role, while the actual charging is driven by coherent dynamics or external resources.
	
	From the perspective of ergotropy, this raises an important question: can a time-independent dissipative mechanism alone steer a tight-binding lattice toward highly excited band states and yield a highly charged steady-state? For a fixed Hamiltonian, maximal ergotropy is achieved when the state is concentrated near the top of the energy spectrum, whereas conventional thermalization yields passive states with vanishing ergotropy~\cite{Farina2018ChargerMediated,Carrega2020Dissipative,Santos2020SelfDischarging,Zakavati2020BoundsPower,Tirone2023NoisyBatteriesCoherence}. Achieving steady states near the upper edge of a tight-binding band therefore requires genuinely non-thermal dissipative processes that break detailed balance in a controlled manner.
	
	In this work, we investigate bond dissipation as an autonomous charging mechanism for lattice quantum batteries. Focusing on tight-binding systems, we demonstrate this mechanism explicitly in a one-dimensional tight-binding chain and in a two-dimensional graphene lattice. We show that appropriately engineered bond-dissipative Lindblad operators can drive the system toward highly excited band states, resulting in substantial steady-state ergotropy without coherent driving. We further find that disorder enhances the charging power, indicating that dissipation-assisted localization effects can be beneficial for energy storage. Moreover, the bond-dissipative charging process remains robust against additional local dephasing, highlighting its resilience under realistic noise conditions. These results identify bond dissipation as a viable and experimentally relevant route for charging quantum batteries in lattice and solid-state platforms.
	
	\section{Theoretical Framework for Open-System Quantum Battery Charging}
	\label{framework}
	
	The theoretical foundation for analyzing energy storage in quantum systems begins by defining the quantum battery as a finite-dimensional system described by a Hamiltonian $H_B$ with spectral decomposition
	\begin{equation}
		H_B = \sum_{n=1}^D E_n |E_n\rangle\langle E_n|,
	\end{equation}
	where $D$ is the Hilbert-space dimension and the eigenvalues are ordered increasingly, $E_{n+1} \ge E_n$. The central quantity of interest is the ergotropy, defined as the maximum work extractable by cyclic unitary operations~\cite{Allahverdyan2004},
	\begin{equation}
		\mathcal{E}(\rho) = \operatorname{Tr}[H_B \rho] - \min_U\operatorname{Tr}[H_B U\rho U^\dagger],
	\end{equation}
	where $\rho$ is the system density matrix and the minimization runs over all unitaries on the system Hilbert space. The minimum of $\operatorname{Tr}[H_B U\rho U^\dagger]$ defines the associated passive state $\sigma_{\rho}$, which is diagonal in the energy eigenbasis,
	\begin{equation}
		\sigma_\rho=\sum_n r_n |E_n\rangle\langle E_n|,
	\end{equation}
	with populations arranged in nonincreasing order, $r_{n+1}\leq r_n$, as the energy increases. Consequently, $\mathcal{E}(\rho) = \operatorname{Tr}[H_B (\rho - \sigma_\rho)]$. Recent work by Malavazi et al.~\cite{Malavazi2025ChargePreserving} has introduced the concepts of isoergotropic states and ergotropy-preserving operations, providing a systematic framework for understanding how stored energy can be manipulated and protected in quantum batteries. Beyond ergotropy, the fluctuations in extractable work also carry operational significance. Recent work by Sarkar et al.~\cite{Sarkar2025Fluctuation} has shown that open-system dynamics offer superior control over such fluctuations compared to closed unitary processes, revealing a fundamental trade-off between resource cost and energy extraction reliability.
	
	In a closed-system charging protocol, the battery evolves under a time-dependent Hamiltonian $H(\tau) = H_B + V(\tau)$, where $V(\tau)$ is a control potential applied over a finite interval $[0,\tau']$. The resulting unitary evolution
	\begin{equation}
		U(\tau') = \mathcal{T}_+ \exp\!\left[-i\int_0^{\tau'} H(s)\, ds\right],
	\end{equation}
	which transforms the initial state $\rho$ into $\rho(\tau') = U(\tau')\rho U^\dagger(\tau')$. The work extracted in such a cycle is $W = \operatorname{Tr}[H_B(\rho - \rho(\tau'))]$, and the ergotropy is obtained by optimizing over admissible controls $V(\tau)$, equivalently over unitaries $U$.
	
	Beyond this operational definition, thermodynamic principles impose universal constraints on extractable work. For a system with Hamiltonian $H_B$ in state $\rho$, an upper bound follows from the free-energy variational principle. Consider the free energy $F(\rho)=\operatorname{Tr}[H_B\rho]-TS(\rho)$, where $S(\rho)=-\operatorname{Tr}[\rho\ln\rho]$ is the von Neumann entropy and $T=\beta^{-1}$ is the temperature. For a given $\rho$, there exists a unique inverse temperature $\bar{\beta}$ such that the Gibbs state $\rho_{\bar{\beta}}$ satisfies $S(\rho_{\bar{\beta}})=S(\rho)$. By the variational principle, $\rho_{\bar{\beta}}=e^{-\bar{\beta}H_B}/\mathcal{Z}$ minimizes $F$ at fixed entropy, implying $F(\rho_{\bar{\beta}})\le F(\rho)$. Using $S(\rho_{\bar{\beta}})=S(\rho)$, this yields
	\begin{equation}
		\operatorname{Tr}[H_B \rho] - \operatorname{Tr}[H_B\rho_{\bar{\beta}}] \ge 0,
	\end{equation}
	and motivates the bound
	\begin{equation}
		W_{\text{bound}} := \operatorname{Tr}[H_B \rho]-\operatorname{Tr}[H_B\rho_{\bar{\beta}}].
		\label{W_bpound}
	\end{equation}
	
	For a single battery in state $\rho$, the maximum work extractable by unitary operations is the ergotropy $\mathcal{E}(\rho)=\operatorname{Tr}[H_B(\rho-\sigma_\rho)]$. In general, $\mathcal{E}(\rho)$ does not reach $W_{\text{bound}}$; the bound becomes asymptotically attainable when many independent copies are available and global operations (e.g., entangling unitaries) are permitted~\cite{Fannes2013quantumbatteries}.
	
	In open-system charging scenarios, the battery interacts with an environment. A paradigmatic route to engineered dissipation is based on repeated interactions with auxiliary systems prepared in thermal states. In this collisional framework, the battery interacts sequentially (each interaction lasting for a time $\tau$) with identical ancillas in state $\omega_\beta(H_A)$. The evolution over one collision is described by the completely positive map $\mathcal{E}(\rho)=\operatorname{Tr}_A[U(\rho\otimes\omega_\beta(H_A))U^\dagger]$, where $U=e^{-i\tau(H_B+H_A+V)}$. If the unitary satisfies a conservation law $[U,H_0+H_A]=0$ for some operator $H_0$ on the battery, the map admits an equilibrium fixed point $\pi=e^{-\beta H_0}/\operatorname{Tr}(e^{-\beta H_0})$. When $H_0\neq H_B$, this steady-state can be active (non-passive) with respect to $H_B$. A canonical example is $H_0=-H_B$, which yields a population-inverted steady-state $\pi\propto e^{\beta H_B}$ and thus finite ergotropy. This illustrates how tailored dissipation can maintain a charged state without time-dependent driving, with the energetic cost encoded in the preparation of the ancillas.
	
	We now specialize to a standard Markovian description of dissipative charging. We consider a composite system described by the total Hamiltonian
	\begin{equation}
		H_{\text{tot}} = H_B + H_R + H_{BR},
	\end{equation}
	where $H_R$ is the reservoir Hamiltonian and $H_{BR}$ denotes the battery--reservoir coupling. Under weak coupling and the Born--Markov approximation~\cite{Moy1999,Breuer2002}, the reduced dynamics of the battery density matrix $\rho(\tau)$ is governed by a Lindblad master equation~\cite{Lindblad1,Lindblad2},
	\begin{align}
		\frac{d \rho(\tau)}{d\tau} = \mathscr{L}[\rho(\tau)]
		&=-i[H_B, \rho(\tau)] \nonumber \\
		&+ \sum_k \gamma_k \left( L_k \rho(\tau) L_k^\dagger - \frac{1}{2}\{L_k^\dagger L_k, \rho(\tau)\} \right),
	\end{align}
	where the jump operators $L_k$ and rates $\gamma_k>0$ model dissipative exchange with the environment, and $\mathscr{L}$ is the Liouvillian superoperator. The formal solution is $\rho(\tau)=e^{\mathscr{L}\tau}\rho(0)$. For generic parameters, the system relaxes to a steady-state $\rho_{\mathrm{ss}}=\lim_{\tau\to\infty}\rho(\tau)$, which corresponds to the right eigenmatrix of $\mathscr{L}$ with zero eigenvalue. The properties of $\rho_{\mathrm{ss}}$ depend on the choice of jump operators: conventional thermal reservoirs drive the system to a passive Gibbs state and thus to self-discharge.
	
	Recent work has shown that dissipation can also be engineered to stabilize nonthermal states, including localized configurations~\cite{Yusipov17} and dissipative transitions between extended and localized phases in disordered or quasiperiodic systems~\cite{WYC_PRL,PL_3D,jiang_Wannier,feng2025localization,roy2025aperiodic} as well as disorder-free localization settings~\cite{Xu_FlatBand,yang2025_Z2}. Moreover, suitably designed dissipators can stabilize selected Hamiltonian eigenstates, yielding nonthermal steady states such as many-body localized phases~\cite{Yusipov18,WYC_MBL} or quantum many-body scars~\cite{Diss_scar}. These observations motivate the expectation that engineered dissipation can steer the system toward a non-passive steady-state dominated by highly excited eigenstates, thereby enabling autonomous charging of quantum batteries.

	\section{Charging in Tight-Binding Lattice Models}
	\label{sec:models}
	
	To demonstrate the proposed dissipative charging mechanism, we consider quantum batteries described by tight-binding Hamiltonians on one- and two-dimensional lattices. The battery Hamiltonian is
	\begin{equation}
		H_B = \sum_{\langle i,j\rangle} \Big(t_{ij} \, c_i^\dagger c_j + \text{H.c.}\Big),
	\end{equation}
	where $c_i^\dagger$ ($c_i$) creates (annihilates) a particle on site $i$, $t_{ij}$ denotes the nearest-neighbor hopping amplitude on bond $\langle i,j\rangle$, and we set $\hbar = 1$. In the absence of interactions and disorder, the single-particle spectrum of $H_B$ forms Bloch bands $E(\mathbf{k})$ labeled by the crystal momentum $\mathbf{k}$. Maximal ergotropy corresponds to populating states near the upper edge of the spectrum, whereas thermal equilibrium favors the bottom of the lowest band.
	
	We engineer a charging process by coupling the lattice to a structured environment via bond dissipation. Specifically, on each bond $\langle i,j\rangle$ we introduce a jump operator
	\begin{equation}
		L_{ij} = (c_i^\dagger + \eta \, c_j^\dagger)(c_i - \eta \, c_j),
		\label{eq:Lij}
	\end{equation}
	with $\eta = e^{i\phi}$ ($\phi \in \mathbb{R}$). This form is motivated by its ability to selectively favor particular phase relations between neighboring sites. When $\eta = 1$, $L_{ij}$ converts an antisymmetric bond mode into a symmetric one, thereby favoring Bloch states with nearly uniform phase along the bond. For $\eta = -1$, the opposite occurs, stabilizing antisymmetric modes.
	
	We first analyze a one-dimensional (1D) homogeneous chain with $L$ sites, periodic boundary conditions, and uniform hopping $t_{ij} = t$. The single-particle eigenvalues are $E_{n} = 2t\cos (k_{n})$ with $k_{n} = \frac{2\pi n}{L}$ for $n = 0,1,\ldots ,L - 1$. The band spans $[-2t, 2t]$, and the states at $k_{n} = 0$ ($k_{n} = \pi$) lie at the band top (bottom). In this work, we set $t=1$.  
	
	To quantify the charging performance, we compute the steady-state ergotropy $\mathcal{E}_{\text{ss}}$ from the steady-state density matrix $\rho_{\text{ss}}$. We obtain $\rho_{\text{ss}}$ by exact diagonalization of the Liouvillian superoperator $\mathcal{L}$ and selecting the (right) eigenmatrix associated with the zero eigenvalue, i.e., $\mathcal{L}[\rho_{\text{ss}}]=0$. Writing $\rho_{\text{ss}}$ in spectral form $\rho_{\text{ss}} = \sum_\alpha p_\alpha |\psi_\alpha\rangle\langle\psi_\alpha|$ with $p_1 \ge p_2 \ge \dots$, and ordering the energies $E_1 \le E_2 \le \dots$, the ergotropy can be expressed as
	\begin{equation}
		\mathcal{E}_{\text{ss}} = \sum_{\alpha,\beta} (E_\beta - E_\alpha) \, p_\alpha \, |\langle \psi_\alpha | \chi_\beta \rangle|^2,
	\end{equation}
	where $|\chi_\beta\rangle$ are the eigenstates of $H_B$ with energies $E_\beta$. Equivalently, to obtain $\mathcal{E}_{\mathrm{ss}}$ numerically we diagonalize $\rho_{\mathrm{ss}}=\sum_\alpha p_\alpha \ket{\psi_\alpha}\!\bra{\psi_\alpha}$ with $p_1\ge p_2\ge\cdots$, and $H_B=\sum_\beta E_\beta \ket{\chi_\beta}\!\bra{\chi_\beta}$ with $E_1\le E_2\le\cdots$. The passive state associated with $\rho_{\mathrm{ss}}$ is then constructed by assigning the largest populations to the lowest energy levels,
	\begin{equation}
		\sigma_{\mathrm{ss}}=\sum_{\beta} p_\beta \ket{\chi_\beta}\!\bra{\chi_\beta},
	\end{equation}
	which yields $\mathcal{E}_{\mathrm{ss}}=\operatorname{Tr}\!\left[H_B(\rho_{\mathrm{ss}}-\sigma_{\mathrm{ss}})\right]$.
	
\begin{figure}[!ht]
		\centering
		\includegraphics[width=1\linewidth]{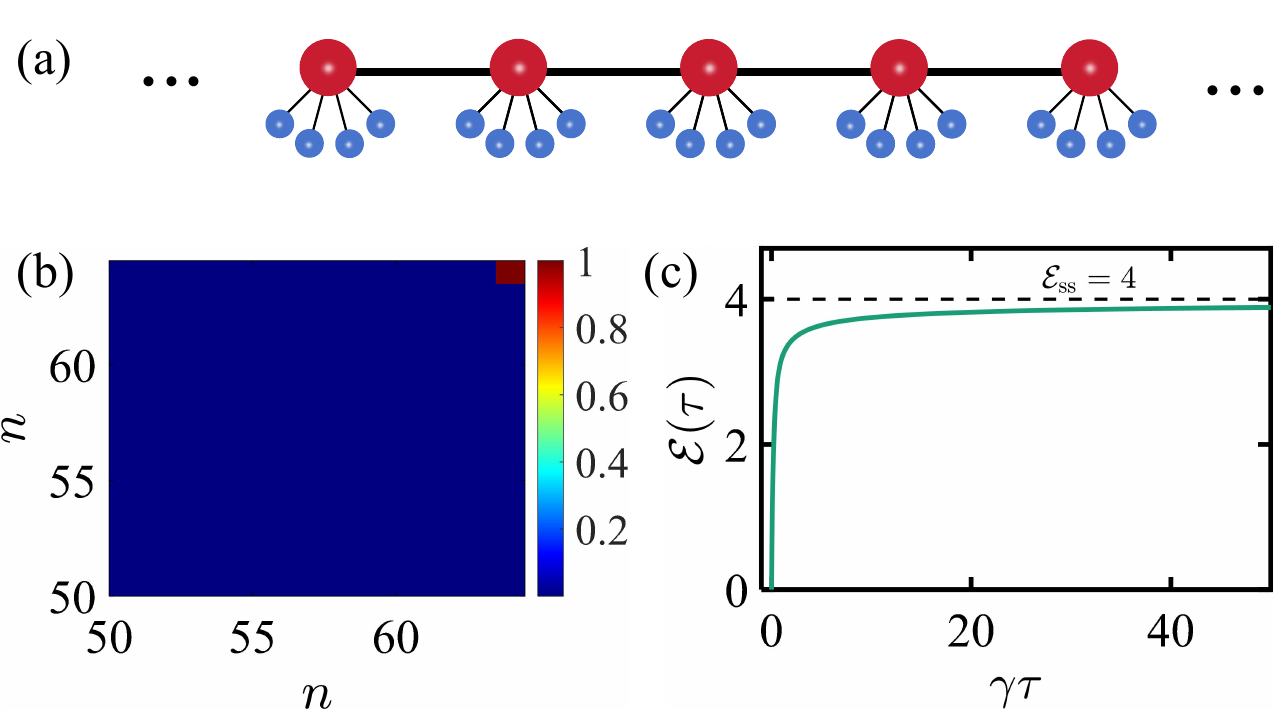}
		\caption{Clean tight-binding quantum battery. (a) Schematic of a dissipative tight-binding chain. Red (blue) spheres denote lattice sites (local baths). (b) Energy spectrum and eigenstate distribution for a one-dimensional tight-binding chain with periodic boundary conditions. The steady-state occupation (color intensity) under bond dissipation with $\eta=1$ is concentrated near the top of the band ($k \approx 0$), corresponding to large ergotropy. (c) Time evolution of the extractable work $\mathcal{E}(\tau)/\mathcal{E}_{\text{ss}}$ starting from the passive state associated with the steady-state. The ergotropy saturates within a few inverse dissipation times $1/\gamma$. Unless stated otherwise, numerical results are obtained for $L=64$ with periodic boundary conditions.}
		\label{Fig1}
	\end{figure}

	When $\eta = 1$, the dissipator suppresses out-of-phase bond modes ($k \approx \pi$) and pumps population into single-particle states with a nearly in-phase profile between neighboring sites ($k \approx 0$). Thus, the dissipation prepares a steady-state concentrated near the band top, as illustrated in Fig.~\ref{Fig1}(a). Starting from the passive state associated with $\rho_{\text{ss}}$, the ergotropy
	\begin{equation}
		\mathcal{E}(\tau)=\operatorname{Tr}[H_B \rho(\tau)] - \min_U\operatorname{Tr}[H_B U\rho(\tau) U^\dagger],
	\end{equation}
	grows from zero and saturates to its steady-state value on a timescale set by $1/\gamma$, as shown in Fig.~\ref{Fig1}(b). We quantify the charging speed by the average power over the interval in which the ergotropy increases from zero to $99\%$ of its saturation value
	\begin{equation}
		P=\frac{\mathcal{E}(\tau_{0.99})}{\tau_{0.99}},
		\label{Power_99}
	\end{equation}
	where $\tau_{0.99}$ is the time at which the ergotropy reaches $\mathcal{E}(\tau_{0.99})=0.99\mathcal{E}_{\mathrm{ss}}$.
	
\begin{figure}[!ht]
		\centering
		\includegraphics[width=1\linewidth]{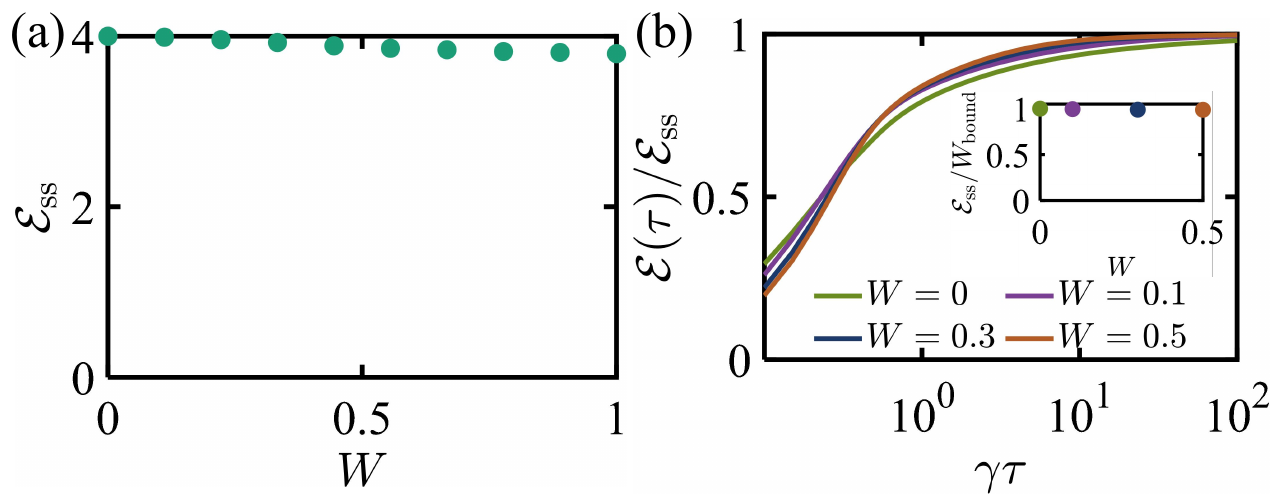}
		\caption{
			Effect of on-site disorder on charging performance in a one-dimensional chain. (a) Steady-state ergotropy $\mathcal{E}_{\text{ss}}$ as a function of disorder strength $W$. (b) Charging dynamics $\mathcal{E}(\tau)/\mathcal{E}_{\text{ss}}$ for different disorder strengths. The corresponding average powers, defined in Eq.~\eqref{Power_99}, are $P_{W=0}=0.15$, $P_{W=0.1}=0.19$, $P_{W=0.3}=0.25$, and $P_{W=0.5}=0.34$. The inset shows that the amount of work extractable from the highly excited charged state via the simple unitary operation defined in this work is close to the bound imposed by the second law of thermodynamics. All data in this figure are computed using periodic boundary conditions.}
		\label{Fig2}
	\end{figure}

	The maximal ergotropy in the clean chain relies on translational symmetry. We now show that the mechanism persists in the presence of on-site disorder. To model disorder, we introduce an on-site potential $V_{\text{dis}} = \sum_i \epsilon_i c_i^\dagger c_i$, where the random energies $\epsilon_i$ are independently drawn from a uniform distribution over $[-W/2, W/2]$, with $W$ denoting the disorder strength. Disorder modifies the spectrum and localizes single-particle eigenstates, yet the engineered bond dissipation continues to pump population into the highest available energy modes. Consequently, while $\mathcal{E}_{\text{ss}}$ is reduced relative to the clean case, the stored work remains substantial for moderate disorder, as shown in Fig.~\ref{Fig2}(a).
	
	Solving for the steady-state confirms that $\rho_{\text{ss}}$ is predominantly supported by eigenstates with positive energies. Moreover, by studying the full dissipative dynamics we find that disorder enhances the charging power. Although the final ergotropy decreases slightly, the approach to $\rho_{\text{ss}}$ is faster, as shown in Fig.~\ref{Fig2}(b). This acceleration is consistent with disorder-induced localization, which suppresses low-energy transport channels and effectively funnels dissipation into the high-energy sector. Interestingly, this phenomenon aligns with the observation in Ref.~\cite{Gyhm2024Beneficial} that charging power is not a monotonic function of entanglement; instead, the precise quantum advantage depends on the interplay between the quantum state and the charging Hamiltonian. Furthermore, from the perspective of fluctuation control, open-system dynamics as studied here generically offer better control over fluctuations in extractable energy compared to closed unitary processes~\cite{Sarkar2025Fluctuation}.
	
	In Fig.~\ref{Fig2}(b), we compare the steady-state ergotropy ${\cal E}_{\rm ss}$ with the upper bound $W_{\text{bound}}$ obtained from the free-energy variational principle [Eq.~\eqref{W_bpound}]. We find that ${\cal E}_{\rm ss}$ remains close to $W_{\text{bound}}$ for moderate disorder, indicating that the dissipative steady-state is nearly optimal without requiring multi-copy global operations.
	
\begin{figure}[!ht]
		\centering
		\includegraphics[width=1\linewidth]{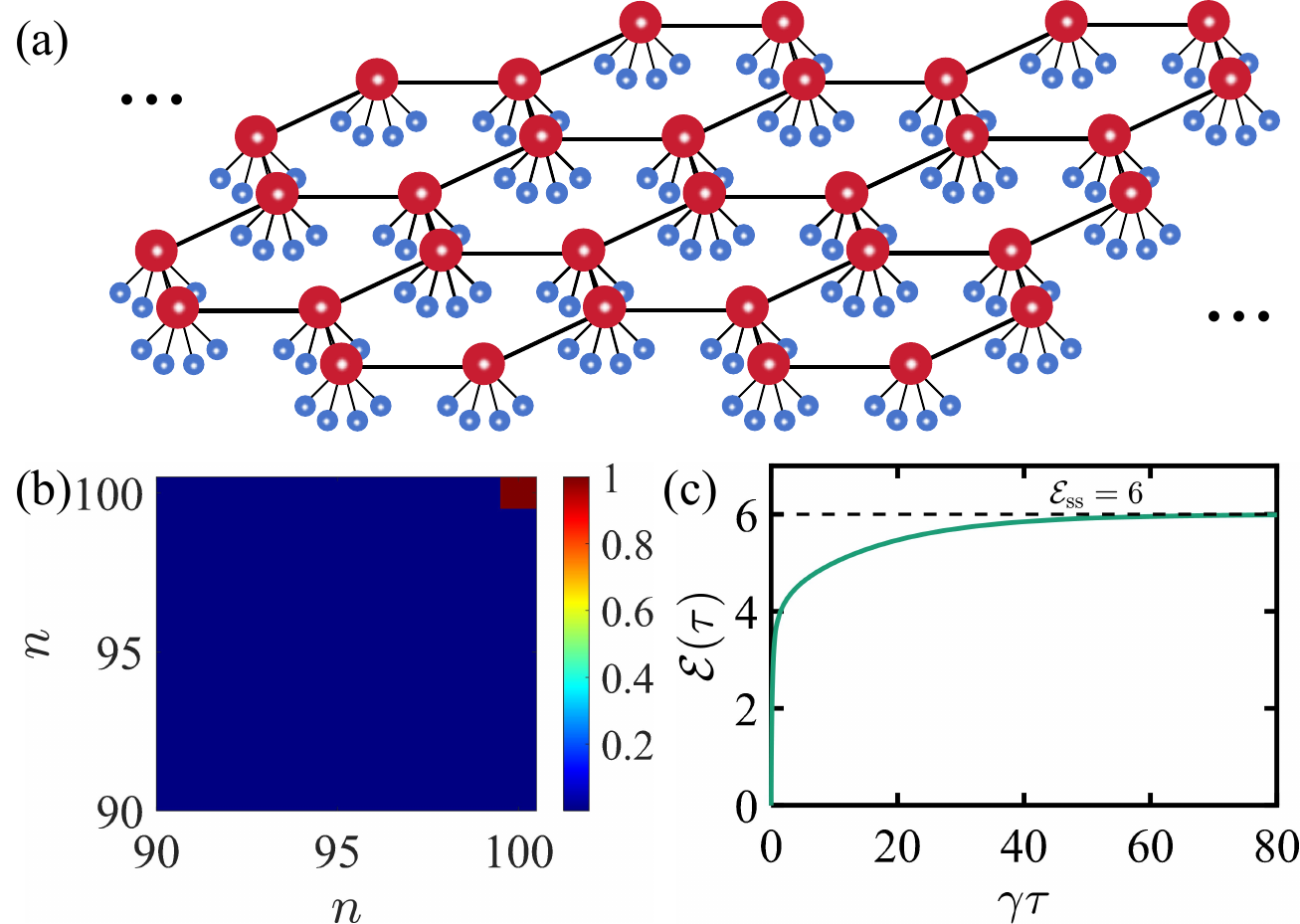}
		\caption{
			Two-dimensional graphene lattice as a quantum battery. (a) Schematic of dissipative graphene. Red (blue) spheres denote lattice sites (local baths). (b) Steady-state occupation on eigenstates under bond dissipation with $\eta=1$, showing preferential population of high-energy states. (c) Temporal evolution of $\mathcal{E}(\tau)$ in the graphene lattice, starting from the passive state associated with the steady-state. All data in this figure are computed using periodic boundary conditions.}
		\label{Fig3}
	\end{figure}

	Next we consider a two-dimensional hexagonal lattice that describes the electronic structure of graphene as illustrated in Fig.~\ref{Fig3}(a). Applying the same bond dissipation \eqref{eq:Lij} with $\eta = 1$ uniformly across all nearest-neighbor bonds drives the system into a steady-state predominantly populated by high-energy eigenstates near the top of the upper band, as shown in Fig.~\ref{Fig3}(b). The corresponding steady-state ergotropy $\mathcal{E}_{\text{ss}}$ is therefore large, demonstrating that the dissipative charging mechanism extends to higher-dimensional geometries.
	
	We also examine the charging dynamics in the graphene lattice. Starting from the passive state associated with the steady-state, the extractable work $\mathcal{E}(\tau)$ increases monotonically and saturates to its steady-state value within a few characteristic times $1/\gamma$, as depicted in Fig.~\ref{Fig3}(c), indicating efficient charging in two dimensions.
	
	\begin{figure}[!ht]
		\centering
		\includegraphics[width=1\linewidth]{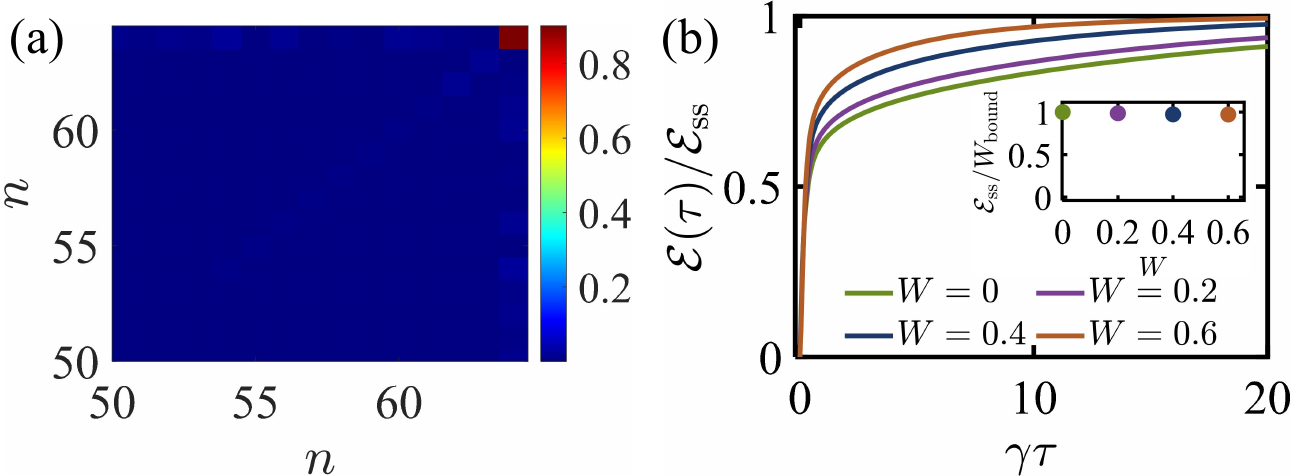}
		\caption{
			Robustness of charging in a disordered graphene lattice. (a) Steady-state occupation on eigenstates under bond dissipation ($\eta=1$) with finite disordered potential. (b) Comparison of charging dynamics $\mathcal{E}(\tau)/\mathcal{E}_{\text{ss}}$ with and without disorder. The corresponding average powers [Eq.~\eqref{Power_99}] are $P_{W=0}=0.14$, $P_{W=0.2}=0.16$, $P_{W=0.4}=0.24$, and $P_{W=0.6}=0.38$. The inset shows $\mathcal{E}_{\text{ss}}$ versus disorder strength $W$. All data in this figure are computed using periodic boundary conditions.}
		\label{Fig4}
	\end{figure}
	
	To examine robustness in two dimensions, we introduce on-site disorder by adding a random potential $H_{\text{dis}}$ as in the one-dimensional case. We find that the steady-state remains dominated by high-energy eigenstates for moderate disorder, retaining a substantial ergotropy [Fig.~\ref{Fig4}(a)]. Moreover, disorder enhances the charging power: although the final ergotropy decreases slightly, the rate at which ergotropy accumulates increases and the battery reaches $\rho_{\text{ss}}$ more rapidly [Fig.~\ref{Fig4}(b)]. We also compare ${\cal E}_{\rm ss}$ with $W_{\text{bound}}$ and find that ${\cal E}_{\rm ss}$ remains a sizable fraction of the bound, indicating near-optimal performance for a single lattice system. This behavior is consistent with disorder-induced localization suppressing low-energy channels and funneling population into the high-energy sector.

	\section{Experimental realization for bond dissipation and robustness to local dephasing}
	
\begin{figure}[h]
		\centering
		\includegraphics[width=1\linewidth]{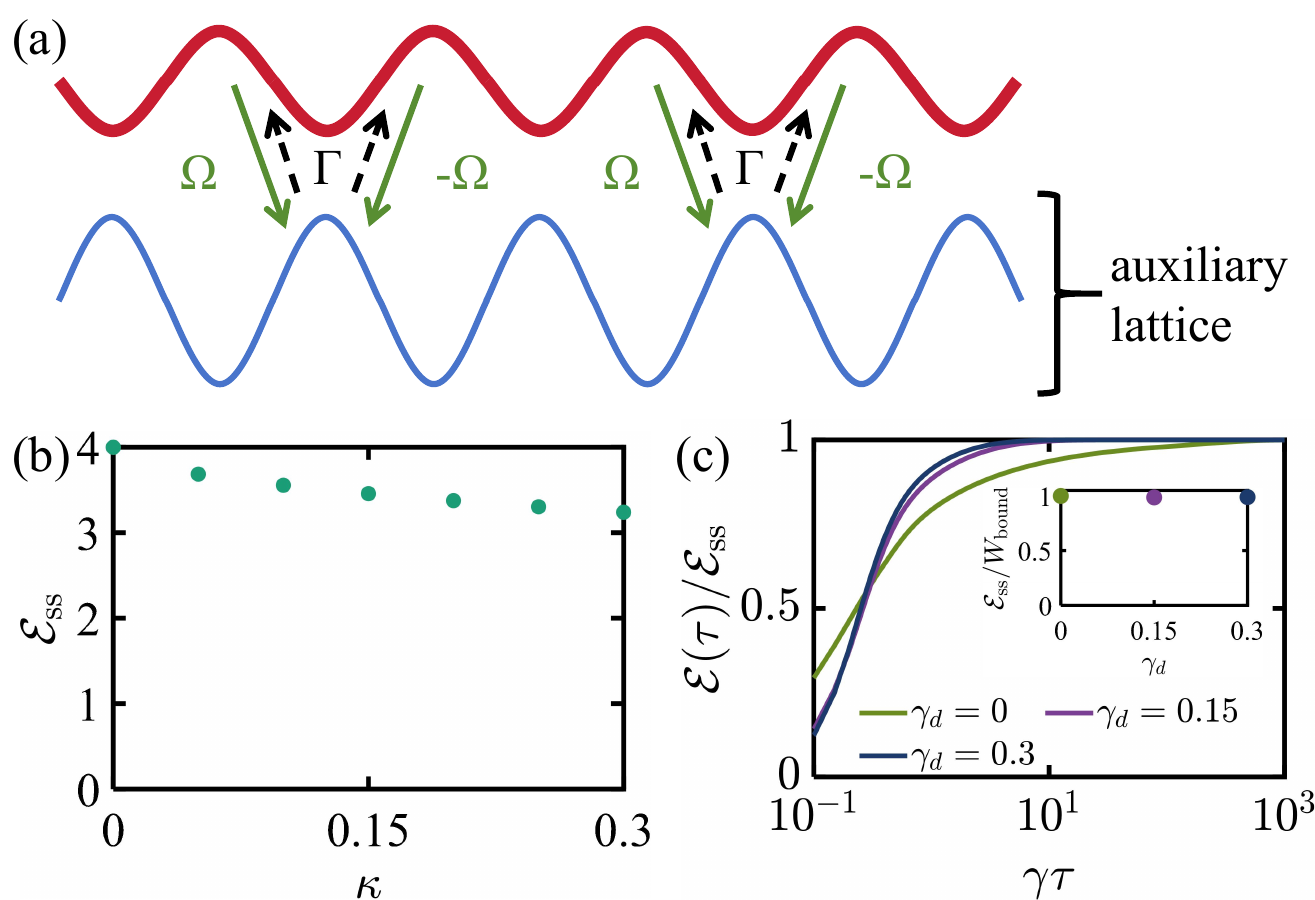}
		\caption{Dissipative graphene and the effects of dephasing. (a) Schematic for implementing local dissipation operators. The lower and upper lattices correspond to the auxiliary system and the battery, respectively. Two adjacent sites in the battery are coupled to an intermediate auxiliary site via Raman lasers with opposite Rabi amplitudes $\pm\Omega$, realizing the annihilation process $(c_i - c_{i+1})$. The subsequent creation process $(c_i^\dagger + c_{i+1}^\dagger)$ is achieved through isotropic spontaneous decay at rate $\Gamma$ back to the original sites. This combined excitation--decay cycle effectively implements the dissipator in Eq.~\eqref{eq:Lij}. (b) Steady-state ergotropy under different dephasing strengths. (c) Comparison of charging dynamics $\mathcal{E}(\tau)/\mathcal{E}_{\text{ss}}$ with and without dephasing. The powers for three different dephasing strengths are: $P_{\gamma_d=0}=0.15$, $P_{\gamma_d=0.15}=0.67$, and $P_{\gamma_d=0.3}=1.21$. Moderate dephasing accelerates the approach to the steady-state. The inset shows that the amount of work extractable from the highly excited charged state via the simple unitary operation defined in this work is very close to the bound imposed by the second law of thermodynamics. All data in this figure are computed using periodic boundary conditions.}
		\label{Fig_setup}
	\end{figure}

	The implementation of local dissipators offers significant advantages over nonlocal operators in terms of experimental feasibility. We present a concrete scheme for realizing the bond-dissipative operators in Eq.~\eqref{eq:Lij}; a detailed derivation of the effective Lindblad dynamics and the associated dissipation strength is provided in Appendix~\ref{app:derivation}. The basic architecture employs two parallel optical lattices: a primary lattice hosting the battery system and an auxiliary lattice that mediates the engineered dissipation, as sketched in Fig.~\ref{Fig_setup}(a).
	
	The annihilation component of the jump operator, $(c_i - e^{i\phi}c_j)$, is implemented by coherently coupling two neighboring sites of the system lattice to a shared auxiliary site located between them. This coupling can be induced via Raman lasers with opposite amplitudes $\Omega$ and $-\Omega$, which selectively excite the antisymmetric superposition of atoms (or spins) on the two system sites. Following this excitation, the creation process $(c_i^\dagger + e^{i\phi}c_j^\dagger)$ is realized through spontaneous emission from the auxiliary site back to the system lattice. Because spontaneous emission is isotropic, it naturally generates a symmetric superposition, thereby completing one cycle of the dissipative transition. The phase parameter $\phi$ can be tuned by adjusting the relative phase of the Raman lasers; alternatively, in superconducting-circuit architectures it can be controlled via the phases of microwave drives coupled to an array of resonators~\cite{Yusipov17,BHchain}.
	
	In experimental settings, quantum systems are exposed to environmental noise, among which local dephasing is a common decoherence mechanism. We therefore assess the robustness of the proposed dissipative charging scheme against dephasing. To this end, we include local dephasing channels described by Lindblad operators $O_j = c_j^\dagger c_j$ with dephasing rate $\gamma_d$. The resulting dynamics is governed by the master equation
	\begin{equation}
		\dot{\rho} = -i[H_B, \rho] + \gamma \sum_{\langle i,j\rangle} \mathcal{D}[L_{ij}](\rho) + \gamma_d \sum_j \mathcal{D}[O_j](\rho),
	\end{equation}
	where $\mathcal{D}[O](\rho) = O\rho O^\dagger - \frac{1}{2}\{O^\dagger O, \rho\}$.
	
	We numerically simulate the charging dynamics of the one-dimensional chain under the joint influence of bond dissipation ($\gamma$) and local dephasing ($\gamma_d$). We find that even when $\gamma_d$ is comparable to $\gamma$, the steady-state ergotropy $\mathcal{E}_{\text{ss}}$ remains a substantial fraction of its value in the clean case, as shown in Fig.~\ref{Fig_setup}(b). Moreover, the time evolution of the extractable work $\mathcal{E}(\tau)$ shows that the charging process remains effective, with the battery reaching its near-steady value within a few inverse dissipation times $1/\gamma$ despite the presence of dephasing [Fig.~\ref{Fig_setup}(c)]. These results indicate that the charging protocol is robust against phase noise and is therefore compatible with noisy solid-state and cold-atom platforms. Beyond charging, the protection of already-stored ergotropy against environmental noise can be further enhanced by techniques such as the weak-measurement protocol proposed in Ref.~\cite{Malavazi2025WeakMeasurement}, which offers a complementary strategy for mitigating energy loss in open quantum batteries.
	
	These considerations confirm that the dissipative charging mechanism is robust against realistic dephasing noise and amenable to implementation in state-of-the-art quantum simulation platforms. The combination of robustness and experimental accessibility positions engineered bond dissipation as a promising route toward practical quantum energy storage devices.
	
	In summary, the bond-dissipative operator \eqref{eq:Lij}, when applied uniformly across a tight-binding lattice, can drive the system autonomously toward a non-passive steady-state rich in high-energy excitations. The mechanism works for both one- and two-dimensional geometries and remains effective in the presence of moderate disorder. This establishes engineered bond dissipation as a versatile and experimentally accessible route for charging quantum batteries in solid-state and cold-atom platforms.
	
	\section{Conclusion and outlook}
	\label{Conclusion}
	
	In this work, we have investigated dissipative charging of lattice quantum batteries via bond dissipation in tight-binding systems. Within a Markovian open-quantum-system framework, we showed that appropriately engineered bond-dissipative Lindblad operators can autonomously steer the system toward highly excited band states, resulting in steady states with substantial ergotropy in the absence of coherent driving.
	Using a one-dimensional tight-binding chain and a two-dimensional graphene lattice as paradigmatic examples, we explicitly demonstrated the charging dynamics and steady-state properties of bond-dissipative quantum batteries. We found that the presence of disorder enhances the charging power, indicating that dissipation-assisted localization effects can play a constructive role in energy storage. Moreover, we showed that the bond-dissipative charging process remains robust against additional local dephasing noise, underscoring its resilience under realistic environmental perturbations.
	
	Our results highlight bond dissipation as a physically transparent and experimentally relevant mechanism for charging lattice quantum batteries. Its feasibility is further supported by recent advances in experimental platforms for open quantum systems \cite{Exp1,Exp2,Exp3,Exp4,Exp5,Exp6,Exp7,Exp8,Exp_new1,Exp_new2,Exp_new3,Ferraro2026Opportunities,Gyhm2024Beneficial}, which have matured the techniques needed to characterize such dissipative protocols. Experimentally, one can monitor the entire charging process, from the initial relaxation to the establishment of the non-passive steady-state amidst decoherence. This allows for the direct extraction of key performance metrics, including the charging efficiency, robustness against noise, and the fundamental limits of the mechanism. More broadly, our work illustrates how engineered dissipation tailored to the band structure of a tight-binding Hamiltonian can be harnessed as a resource, rather than a limitation, in open many-body quantum systems. These findings are directly relevant for solid-state and quantum-simulation platforms where lattice Hamiltonians and controlled dissipation are naturally available, offering a practical pathway toward realizing autonomous quantum energy storage. 
	
	A promising direction for future research is to combine the dissipative charging mechanism studied here with ergotropy-preserving operations~\cite{Malavazi2025ChargePreserving}. Such hybrid protocols could enable both efficient charging and flexible manipulation of stored energy, potentially leading to more versatile quantum battery platforms. Furthermore, the weak-measurement protection scheme~\cite{Malavazi2025WeakMeasurement} could be integrated to enhance the resilience of charged states against decoherence, extending the storage lifetime of the battery. Extending the present analysis to include fluctuations in extractable work, as studied in Ref.~\cite{Sarkar2025Fluctuation}, would also be valuable for characterizing the reliability of the charging protocol.

	\appendix
	\section{Derivation of the Bond-Dissipative Lindblad Operator}
	\label{app:derivation}
	
	In this appendix, we outline a microscopic derivation of the effective Lindblad master equation with local bond dissipators [Eq.~\eqref{eq:Lij}]. The procedure follows established adiabatic-elimination methods in open quantum systems \cite{Jump1,Jump2}.
	
	We start from a microscopic model comprising three parts: the battery system $S$, an array of auxiliary sites $A$, and a structured reservoir $R$. The total Hamiltonian reads
	\begin{equation}
		H_{\text{total}} = H_S + H_A + H_R + H_{SA} + H_{AR},
	\end{equation}
	where $H_S = \sum_{\langle i,j\rangle} t_{ij} c_i^\dagger c_j + \text{H.c.}$ is the battery Hamiltonian. The auxiliary lattice is described by $H_A = \epsilon \sum_m a_m^\dagger a_m$, with $a_m$ the annihilation operator on the $m$th ancillary site and $\epsilon$ its on-site energy. The reservoir is modeled as a bosonic bath $H_R = \sum_{\mathbf{k}} \omega_{\mathbf{k}} b_{\mathbf{k}}^\dagger b_{\mathbf{k}}$.
	
	The system--ancilla coupling is engineered via Raman lasers that link two neighboring system sites $i,j$ to a common ancillary site $m$ between them:
	\begin{equation}
		H_{SA}(\tau) = \sum_m \Big[ \Omega(\tau)\, a_m^\dagger \big( c_i - e^{i\phi}c_j \big) + \text{H.c.} \Big],
	\end{equation}
	where $\Omega(\tau) = \Omega e^{-i\omega_L \tau}$ is the time-dependent Rabi amplitude and $\phi$ is a controllable phase. This coupling selectively excites the antisymmetric superposition on the bond for $\phi=0$. The ancilla--reservoir interaction is taken as
	\begin{equation}
		H_{AR} = \sum_{m,\mathbf{k}} \Big( g_{\mathbf{k}} a_m^\dagger b_{\mathbf{k}} + \text{H.c.} \Big).
	\end{equation}
	
	We first integrate out the reservoir. Assuming that the reservoir is at zero temperature and that its correlation time is short compared to the system dynamics, we trace out the bath degrees of freedom within the Born--Markov approximation. This produces a Lindblad dissipator for the ancilla modes,
	\begin{equation}
		\mathcal{D}_A[\rho_{SA}] = \Gamma \sum_m \Big( 2 a_m \rho_{SA} a_m^\dagger - \{ a_m^\dagger a_m, \rho_{SA} \} \Big),
	\end{equation}
	where the decay rate $\Gamma = 2\pi \sum_{\mathbf{k}} |g_{\mathbf{k}}|^2 \delta(\epsilon - \omega_{\mathbf{k}})$ is determined by the bath spectral density at the ancilla frequency $\epsilon$.
	
	In the limit of large laser detuning $\Delta = \omega_L - \epsilon$ and fast ancilla decay $\Gamma \gg |\Omega|$, we further eliminate the ancillary modes. Setting the Heisenberg equation for $a_m$ to zero in the mean-field approximation gives
	\begin{equation}
		0 \approx -i\Delta \, a_m - i\Omega\big(c_i - e^{i\phi}c_j\big) - \frac{\Gamma}{2} a_m,
	\end{equation}
	which yields
	\begin{equation}
		a_m \approx -\frac{\Omega}{\Delta + i\Gamma/2} \big(c_i - e^{i\phi}c_j\big).
	\end{equation}
	For $|\Delta| \gg \Gamma$, this simplifies to $a_m \approx -(\Omega/\Delta)\big(c_i - e^{i\phi}c_j\big)$.
	
	Substituting this relation into the ancilla dissipator $\mathcal{D}_A$ and using that the decay back to the system from the ancilla is isotropic (i.e., symmetric with respect to the two system sites), we obtain the effective jump operator for the bond $\langle i,j\rangle$:
	\begin{equation}
		L_{ij} \propto \big(c_i^\dagger + e^{-i\phi}c_j^\dagger\big)\big(c_i - e^{i\phi}c_j\big).
	\end{equation}
	The Hermitian conjugate $L_{ij}^\dagger$ corresponds to the reverse process. After normalization, we recover the form used in the main text:
	\begin{equation}
		L_{ij} = \big(c_i^\dagger + \eta c_j^\dagger\big)\big(c_i - \eta c_j\big), \qquad \eta = e^{i\phi}.
	\end{equation}
	
	The overall strength $\gamma$ of the engineered dissipation is given by the product of the excitation probability, namely $|\Omega/\Delta|^2$, and the ancilla decay rate $\Gamma$:
	\begin{equation}
		\gamma = \Gamma \, \frac{|\Omega|^2}{\Delta^2 + (\Gamma/2)^2} \approx \Gamma \frac{|\Omega|^2}{\Delta^2} \quad (\text{for } |\Delta| \gg \Gamma).
	\end{equation}
	
	Typical experimental parameters in cold-atom setups \cite{Jump2,Levine2019} give $\Omega \sim 2\pi \times 1\;\text{MHz}$, $\Gamma \sim 2\pi \times 10\;\text{MHz}$, and $\Delta \sim 2\pi \times 50\;\text{MHz}$, leading to $\gamma \sim 0.01\Omega - 0.1\Omega$ (in units of $\hbar=1$), with Rabi frequencies $\Omega \approx 2\pi \times 1\;\text{MHz}$ \cite{Rydberg_Exp1,Rydberg_Exp2,Levine2019}. This demonstrates the feasibility of implementing the proposed bond-dissipative charging protocol with existing quantum simulation platforms.
	
	\section*{Acknowledgements}
	The work is supported by the National Natural Science Foundation of China (Grant No. 12304290 and No. 12505017), and Beijing National Laboratory for Condensed Matter Physics (2025BNLCMPKF017). LP also acknowledges support from the Fundamental Research Funds for the Central Universities. \\
	
	\bibliography{References}
	
\end{document}